\documentclass[useAMS,usenatbib]{mn2e}
\usepackage{times}
\usepackage{xspace}
\usepackage{graphicx}
\usepackage{tabularx}
\newcommand{\ignore}[1]{}
\providecommand{\ao}{}
\renewcommand{\ao}{adaptive optics (AO)\renewcommand{\ao}{AO\xspace}\renewcommand{\Ao}{AO\xspace}\xspace}
\newcommand{\Ao}{Adaptive optics (AO)\renewcommand{\ao}{AO\xspace}\renewcommand{\Ao}{AO\xspace}\xspace}
\newcommand{\wfs}{wavefront sensor (WFS)\renewcommand{\wfs}{WFS\xspace}\renewcommand{\wfss}{WFSs\xspace}\xspace}
\newcommand{\wfss}{wavefront sensors (WFSs)\renewcommand{\wfs}{WFS\xspace}\renewcommand{\wfss}{WFSs\xspace}\xspace}
\newcommand{\shwfs}{Shack-Hartmann \wfs (SHWFS)\renewcommand{\shwfs}{SHWFS\xspace}\xspace}
\newcommand{\dm}{deformable mirror (DM)\renewcommand{\dm}{DM\xspace}\renewcommand{\dms}{DMs\xspace}\renewcommand{\Dms}{DMs\xspace}\renewcommand{\Dm}{DM\xspace}\xspace}
\newcommand{\dms}{deformable mirrors (DMs)\renewcommand{\dm}{DM\xspace}\renewcommand{\dms}{DMs\xspace}\renewcommand{\Dms}{DMs\xspace}\renewcommand{\Dm}{DM\xspace}\xspace}
\newcommand{\Dms}{Deformable mirrors (DMs)\renewcommand{\dm}{DM\xspace}\renewcommand{\dms}{DMs\xspace}\renewcommand{\Dms}{DMs\xspace}\renewcommand{\Dm}{DM\xspace}\xspace}
\newcommand{\Dm}{Deformable mirror (DM)\renewcommand{\dm}{DM\xspace}\renewcommand{\dms}{DMs\xspace}\renewcommand{\Dms}{DMs\xspace}\renewcommand{\Dm}{DM\xspace}\xspace}

\newcommand{\shs}{Shack-Hartmann sensor (SHS)\renewcommand{\shs}{SHS\xspace}\renewcommand{\shss}{SHSs\xspace}\xspace}
\newcommand{\shss}{Shack-Hartmann sensors (SHSs)\renewcommand{\shs}{SHS\xspace}\renewcommand{\shss}{SHSs\xspace}\xspace}
\newcommand{\lgs}{laser guide star (LGS)\renewcommand{\lgs}{LGS\xspace}\renewcommand{\lgss}{LGSs\xspace}\xspace}
\newcommand{\lgss}{laser guide stars (LGSs)\renewcommand{\lgs}{LGS\xspace}\renewcommand{\lgss}{LGSs\xspace}\xspace}
\newcommand{\ngs}{natural guide star (NGS)\renewcommand{\ngs}{NGS\xspace}\renewcommand{\ngss}{NGSs\xspace}\xspace}
\newcommand{\ngss}{natural guide stars (NGSs)\renewcommand{\ngs}{NGS\xspace}\renewcommand{\ngss}{NGSs\xspace}\xspace}
\newcommand{\mems}{Micro-Electro-Mechanical Systems (MEMS)\renewcommand{\mems}{MEMS\xspace}\xspace}
\newcommand{\snr}{signal to noise ratio (SNR)\renewcommand{\snr}{SNR\xspace}\xspace}
\newcommand{\glao}{ground layer \ao (GLAO)\renewcommand{\glao}{GLAO\xspace}\xspace}
\newcommand{\moao}{multi-object \ao (MOAO)\renewcommand{\moao}{MOAO\xspace}\renewcommand{\Moao}{MOAO\xspace}\xspace}
\newcommand{\Moao}{Multi-object \ao (MOAO)\renewcommand{\moao}{MOAO\xspace}\renewcommand{\Moao}{MOAO\xspace}\xspace}
\newcommand{\mcao}{multi-conjugate adaptive optics (MCAO)\renewcommand{\mcao}{MCAO\xspace}\xspace}
\newcommand{\ltao}{laser tomographic adaptive optics (LTAO)\renewcommand{\ltao}{LTAO\xspace}\xspace}
\newcommand{\cpu}{central processing unit (CPU)\renewcommand{\cpu}{CPU\xspace}\renewcommand{\cpus}{CPUs\xspace}\xspace}
\newcommand{\cpus}{central processing units (CPUs)\renewcommand{\cpu}{CPU\xspace}\renewcommand{\cpus}{CPUs\xspace}\xspace}
\newcommand{\psf}{point spread function (PSF)\renewcommand{\psf}{PSF\xspace}\renewcommand{\psfs}{PSFs\xspace}\xspace}
\newcommand{\psfs}{point spread functions (PSFs)\renewcommand{\psf}{PSF\xspace}\renewcommand{\psfs}{PSFs\xspace}\xspace}
\newcommand{\fpga}{field programmable gate array (FPGA)\renewcommand{\fpga}{FPGA\xspace}\renewcommand{\fpgas}{FPGAs\xspace}\xspace}
\newcommand{\fpgas}{field programmable gate arrays (FPGAs)\renewcommand{\fpga}{FPGA\xspace}\renewcommand{\fpgas}{FPGAs\xspace}\xspace}
\newcommand{\sor}{successive over-relaxation (SOR)\renewcommand{\sor}{SOR\xspace}\xspace}
\newcommand{\fdpcg}{Fourier domain pre-conditioned gradient (FDPCG)\renewcommand{\fdpcg}{FDPCG\xspace}\xspace}
\newcommand{\map}{maximum a-posteriori (MAP)\renewcommand{\map}{MAP\xspace}\xspace}
\newcommand{\elt}{Extremely Large Telescope (ELT)\renewcommand{\elt}{ELT\xspace}\renewcommand{\elts}{ELTs\xspace}\xspace}
\newcommand{\elts}{Extremely Large Telescopes (ELTs)\renewcommand{\elt}{ELT\xspace}\renewcommand{\elts}{ELTs\xspace}\xspace}
\newcommand{\dugall}{Durham University generalised adaptive optics laser laboratory (DUGALL)\renewcommand{\dugall}{DUGALL\xspace}\xspace}
\newcommand{\fwhm}{full-width at half-maximum (FWHM)\renewcommand{\fwhm}{FWHM\xspace}\xspace}
\newcommand{\wht}{William Herschel Telescope (WHT)\renewcommand{\wht}{WHT\xspace}\xspace}
\newcommand{\emccd}{electron multiplying CCD (EMCCD)\renewcommand{\emccd}{EMCCD\xspace}\xspace}
\newcommand{\dasp}{Durham \ao simulation platform (DASP)\renewcommand{\dasp}{DASP\xspace}\xspace}
\newcommand{\eelt}{European \elt (E-ELT)\renewcommand{\eelt}{E-ELT\xspace}\xspace}
\newcommand{\mpi}{Message Passing Interface (MPI)\renewcommand{\mpi}{MPI\xspace}\xspace}
\newcommand{\smp}{symmetric multi-processing (SMP)\renewcommand{\smp}{SMP\xspace}\xspace}
\newcommand{\svd}{singular value decomposition (SVD)\renewcommand{\svd}{SVD\xspace}\xspace}
\newcommand{\gpu}{graphical processing unit (GPU)\renewcommand{\gpu}{GPU\xspace}\renewcommand{\gpus}{GPUs\xspace}\xspace}
\newcommand{\gpus}{graphical processing units (GPUs)\renewcommand{\gpu}{GPU\xspace}\renewcommand{\gpus}{GPUs\xspace}\xspace}
\newcommand{\fft}{fast Fourier transform (FFT)\renewcommand{\fft}{FFT\xspace}\xspace}
\newcommand{\ifu}{integral field unit (IFU)\renewcommand{\ifu}{IFU\xspace}\xspace}
\newcommand{\darc}{the Durham adaptive optics real-time controller (DARC)\renewcommand{\darc}{DARC\xspace}\renewcommand{\Darc}{DARC\xspace}\xspace}
\newcommand{\Darc}{The Durham adaptive optics real-time controller (DARC)\renewcommand{\darc}{DARC\xspace}\renewcommand{\Darc}{DARC\xspace}\xspace}
\newcommand{\cots}{commercial off-the-shelf (COTS)\renewcommand{\cots}{COTS\xspace}\xspace}
\newcommand{\rtcp}{real-time control pipeline (RTCP)\renewcommand{\rtcp}{RTCP\xspace}\xspace}
\newcommand{\rms}{root-mean-square (RMS)\renewcommand{\rms}{RMS\xspace}\xspace}
\newcommand{\sFPDP}{serial Front Panel Data Port (sFPDP)\renewcommand{\sFPDP}{sFPDP\xspace}\xspace}
\newcommand{\wpu}{wavefront processing unit (WPU)\renewcommand{\wpu}{WPU\xspace}\xspace}
\newcommand{\canary}{CANARY\xspace}

\newcommand{\rtcs}{real-time control system (RTCS)\renewcommand{\rtcs}{RTCS\xspace}\xspace}
\newcommand{\ptp}{point-to-point (PTP)\renewcommand{\ptp}{PTP\xspace}\xspace}
\newcommand{\sse}{streaming SIMD extension (SSE)\renewcommand{\sse}{SSE\xspace}\xspace}
\newcommand{\api}{application programming interface (API)\renewcommand{\api}{API\xspace}\xspace}
\newcommand{\corba}{Common Object Request Broker Architecture (CORBA)\renewcommand{\corba}{CORBA\xspace}\xspace}
\newcommand{\lqg}{linear quadratic gaussian (LQG)\renewcommand{\lqg}{LQG\xspace}\xspace}
\newcommand{\scao}{single conjugate adaptive optics (SCAO)\renewcommand{\scao}{SCAO\xspace}\xspace}
\newcommand{\dma}{direct memory access (DMA)\renewcommand{\dma}{DMA\xspace}\xspace}
\newcommand{\xao}{extreme adaptive optics (XAO)\renewcommand{\xao}{XAO\xspace}\xspace}
\newcommand{\vlt}{Very Large Telescope (VLT)\renewcommand{\vlt}{VLT\xspace}\xspace}
\newcommand{\sparta}{Standard Platform for Advanced Real-Time
  Applications (SPARTA)\renewcommand{\sparta}{SPARTA\xspace}\xspace}
\newcommand{\eso}{European Southern Observatory (ESO)\renewcommand{\eso}{ESO\xspace}\xspace}
\newcommand{\eagle}{ELT Adaptive optics for GaLaxy Evolution (EAGLE)\renewcommand{\eagle}{EAGLE\xspace}\xspace}
\newcommand{\epics}{Exo-Planet Imaging Camera and Spectrograph (EPICS)\renewcommand{\epics}{EPICS\xspace}\xspace}
\newcommand{\iir}{infinite impulse response (IIR)\renewcommand{\iir}{IIR\xspace}\xspace}
\newcommand{\gtc}{Gran Telescopio Canarias (GTC)\renewcommand{\gtc}{GTC\xspace}\xspace}

\title[ELT MOAO DM requirements]{Monte-Carlo simulation of ELT scale
  multi-object adaptive optics deformable mirror requirements and
  tolerances} \author[A. G. Basden, N. A. Bharmal, R. M. Myers, S. L. Morris, T. J. Morris]{A. G. Basden$^{1}$\thanks{E-mail:
    a.g.basden@durham.ac.uk (AGB)}, N. A. Bharmal$^1$, R. M. Myers$^1$,
  S. L. Morris$^1$, T. J. Morris$^1$\\ $^{1}$Department of Physics, South Road, Durham,
  DH1 3LE, UK}

\begin{document}
\maketitle

\begin{abstract}
Multi-object adaptive optics (MOAO) has been demonstrated by the
CANARY instrument on the William Herschel Telescope.  However, for
proposed MOAO systems on the next generation Extremely Large
Telescopes, such as EAGLE, many challenges remain.  Here we
investigate requirements that MOAO operation places on deformable
mirrors (DMs) using a full end-to-end Monte-Carlo AO simulation code.
By taking into consideration a prior global ground-layer (GL)
correction, we show that actuator density for the MOAO DMs can be
reduced with little performance loss.  We note that this reduction is
only possible with the addition of a GL DM, whose order is greater
than or equal to that of the original MOAO mirrors.  The addition of a
GL DM of lesser order does not affect system performance (if tip/tilt
star sharpening is ignored).  We also quantify the maximum mechanical
DM stroke requirements (3.5~$\mu$m desired) and provide tolerances for
the DM alignment accuracy, both lateral (to within an eighth of a
sub-aperture) and rotational (to within 0.2$^\circ$).  By presenting
results over a range of laser guide star asterism diameters, we ensure
that these results are equally applicable for laser tomographic AO
systems.  We provide the opportunity for significant cost savings to
be made in the implementation of MOAO systems, resulting from the
lower requirement for DM actuator density.
\end{abstract}

\section{Introduction}
The proposed next generation optical ground-based \elts, with primary
mirror diameters of over 30~m, are currently in the design phase.
These facilities, which will depend on \ao \citep{adaptiveoptics} for
their operation, will provide astronomers with the necessary
resolutions and light collecting areas to probe the universe with
unprecedented sensitivity.  The 39~m \eelt has a suite of planned
instruments, one of which, the proposed \eagle instrument
\citep{2008SPIE.7014E..53Cshort}, uses \moao
\citep{canaryresultsshort} to deliver a high degree of \ao correction
over a wide field of view.  \Moao systems operate in open-loop,
i.e.\ the wavefront sensors do not sense the changes applied to the
\dms.  The \eagle instrument will operate with six \lgss and up to
five \ngss \citep{2010SPIE.7736E..25Rshort}, delivering correction for up to 20 separate science fields
each 1.65~arcsec in diameter, spread across a 10~arcmin field
of view (with the central 5~arcminutes being well corrected), with a
7.3~arcmin technical field. 

The design of any \ao system requires extensive numerical simulation
and modelling of \ao performance so that key design parameters can be
determined, and to ensure that the science goals will be achievable.
The \dasp \citep{basden5} is a Monte-Carlo time-domain code that has
been developed specifically for \elt simulation, including optional
hardware acceleration \citep{basden4,basden6}.  It is an end-to-end
parallelised code including detailed models of telescope and \ao
systems, allowing high-fidelity models to be produced.

The \elt designs include a large \dm (M4, with $85\times85$ actuators
for the \eelt) early in the telescope's optical train
\citep{eelt,tmt}, optically conjugated close to ground level.
Although \moao instruments typically operate in open loop (with the
\dms placed after the \wfs light has been picked off), this telescope
\dm is visible to the \wfss and therefore is operated in closed-loop,
with the \wfss being sensitive to changes in the \dm surface.
Although theoretically not required for an \moao instrument (which has
its own \dms, one for each corrected line of sight), this \dm can be
used to perform a global, \glao correction across the telescope field
of view.  Previous Monte-Carlo based numerical studies for \eagle
\citep{basden8,2008SPIE.7015E..20F} have generally ignored this \dm,
rather assuming an idealised open-loop \dm for each science field.

In this paper, we investigate some of the benefits that are available
to an \moao instrument by making use of this \glao correction, with
considerations paid to the reductions in the required \moao \dm
mechanical stroke capability and also the order of the \moao \dms
(i.e.\ the actuator count).  We also consider the impact of \dm
misalignment on the performance of this \moao system, thus providing
information about acceptable alignment tolerances.  Comparisons of our
simulation results with other codes, both Monte-Carlo and analytic are
also made.

In \S2 we introduce the simulations including parameters that were
used, and the investigations carried out.  In \S3 we present results
and discuss their implications, and in \S4 we draw our
conclusions.

\section{MOAO simulation details}
For the purposes of this paper, we have developed a model of an \moao
instrument using \dasp.  We assume a 42~m diameter telescope with a
central obscuration of 6~m, ignoring effects due to the secondary
support structure.  We have settled on using older parameters for
telescope diameter rather than the current 39~m diameter, so that
these simulation results can easily be compared with previous
simulations performed before downsizing of the \eelt.  The atmosphere
is modelled using a nine layer profile as given in
table~\ref{tab:atmos}, with a 30~m outer scale and a Fried's parameter
(r$_0$) of 13.5~cm at 500~nm corresponding to seeing of 0.8''.  This
atmospheric profile has been chosen to match that used in many
simulations performed at the \eso \citep{miskaltao}.  Phase screens
are sampled with a 3.125~cm spacing.  The simulation consists of six
\lgss arranged in a regular hexagon with each wavefront sensor being a
Shack-Hartmann sensor with $84\times84$ sub-apertures (each 0.5~m in
the pupil plane), each having $16\times16$ detector pixels.  For
simplicity, and so as not to confuse results unnecessarily, we assume
that the tip-tilt signal from the \lgss is valid, or equivalently,
that \ngs tip-til correction is performed perfectly.  This allows us
to focus on tomographic wavefront reconstruction from the \lgss,
without requiring additional parameters to specify \ngs asterisms.  A
real system would in fact ignore the low order signals from the \lgss,
instead using \ngs information to provide these corrections.
Degradation of \ao correction due to tip-tilt in-determination will
depend on the \ngss themselves, both location within the field of view
and magnitude.  We have ignored this consideration here because it is
a study in itself \citep{2008SPIE.7015E..51G}, and as a consequence,
our results are slightly optimistic.

We include both \lgs spot elongation and cone effect in these
simulations and use a centre of gravity centroiding algorithm to
measure local wavefront gradients. We model a Sodium layer with a mean
90~km distance from the telescope, with a Gaussian intensity profile
with a full-width at half-maximum of 10~km.  Our simulations contain
no \ngss so that we can investigate the tomography purely from the
\lgss.  We operate in a high light level regime, with each
sub-aperture receiving $10^6$ photons per frame, and photon shot noise
is included.  The telescope \glao \dm and the individual \moao \dms
have $85\times85$ actuators, unless otherwise stated.  In this paper,
we concentrate on the on-axis science performance, corresponding to
the location furthest from the \lgss, though also present a
performance map across the telescope field of view.  Unless otherwise
stated, results are given for the percentage of ensquared energy
within 75~mas in H-band (wavelength of 1650~nm), which is a key
performance criteria for \eagle.

\begin{table*}
\centering
\begin{minipage}{150mm}
\begin{tabularx}{\linewidth}{Xccccccccc}
C$_n^2$ profile & Layer 1 & Layer 2 & Layer 3 & Layer 4 & Layer 5 &
  Layer 6 & Layer 7 & Layer 8 & Layer 9 \\ \hline
Height / m & 47 & 140 & 281 & 562 & 1125 & 2250 & 4500 & 9000 & 18000\\
C$_n^2$ \% & 52.24 & 2.6 & 4.44 & 11.6 & 9.89 & 2.95 &
5.98 & 4.3 & 6 \\
Speed / ms$^{-1}$ & 4.55 & 12.61 & 12.61 &  8.73 &  8.73 & 14.55 &
24.25 & 38.8 & 20.37\\
Direction / $^\circ$ & 0 &  36 &  72 & 108 & 144 & 180 & 216 & 252 & 288\\
\end{tabularx}
\caption{A table giving the atmospheric layer heights above the
  primary mirror, and corresponding layer strengths used in the
  simulations here, taken from \citet{miskaltao}.}
\label{tab:atmos}
\end{minipage}
\end{table*}

\begin{figure}
\includegraphics[width=\linewidth]{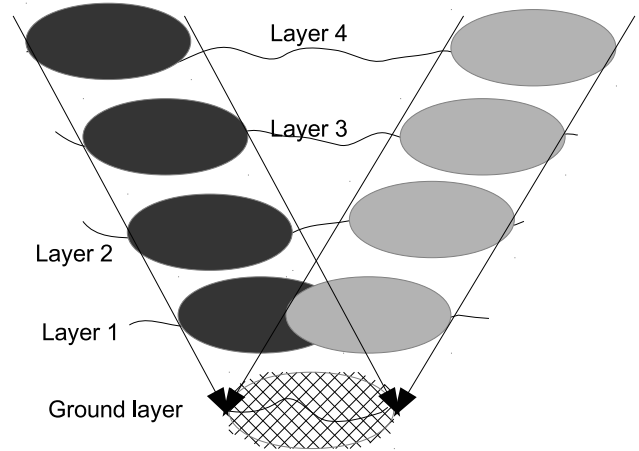}
\caption{A figure showing the separation of GLAO and MOAO DM
  correction.  The GLAO DM (hatched) is used to perform a global
  ground layer correction, while the MOAO DMs (one performing
  correction of light grey areas, one performing correction of dark
  grey areas) only correct higher layers along individual lines of
  sight, and perform no additional ground layer correction.}
\label{fig:layers}
\end{figure}

We use the \glao \dm to perform a global ground layer correction
across the field of view (since this is conjugated to the telescope
pupil).  The \moao \dms are then used to correct only higher layer
turbulence, i.e.\ the \moao \dms are not used for any ground layer
correction.  We make this distinction because the \glao \dm operates
in closed loop while the \moao \dms operate in open-loop (i.e.\ the
\wfss are not sensitive to changes on these \dms).  The \glao and
\moao \dms are therefore correcting independent turbulence with no
interplay between them, as demonstrated in Fig.~\ref{fig:layers}.  In
the cases where the \moao \dm is of a lower order than the \glao \dm,
we realise that the actuators of these two \dms will not be co-aligned
at the ground layer, and thus it might be possible to reduce \dm
fitting error by using the \moao \dm to remove some of the residual
ground layer turbulence corrected by the \glao \dm.  However we do not
do this so as to maintain a clear distinction between the \glao and
\moao corrected turbulence.

The tomographic wavefront reconstruction is performed at the nine
turbulent layers, and the spacing between reconstructed phase points is
dependent on relative layer strength and layer height
\citep{gavelprivate} according to

\begin{equation}
N_\mathrm{i} = \left[ \frac{D_\mathrm{i} r_\mathrm{0}}{D_\mathrm{0} r_\mathrm{i}} \right]^\frac{10}{11} N_\mathrm{0}
\end{equation}
where $N_\mathrm{i}$ is the number of phase points in the i\textsuperscript{th}
layer with $N_\mathrm{0}=85$, $D_\mathrm{i}$ is the diameter of this layer
(which changes with height due to the non-zero field of view), and
$r_\mathrm{i}$ is Fried's parameter for layer $\mathrm{i}$.

We assume an \ao frame rate of 250~Hz, and run simulations for 40~s of
telescope time (10000 iterations) to ensure that the science \psf is
well averaged, which we verify, and use a non-varying r$_0$.  The chosen
frame rate is the baseline for \eagle, and although low, the nature of
open-loop systems mean that \ao system bandwidth is higher than an
equivalent closed-loop system.

Wavefront reconstruction is performed using a regularised
least-squares formulation, based on a sparse Laplacian approximation
of the phase covariance \citep{simScaling}.  Since our light levels
are high and slope measurements almost noiseless, this approximates to
a minimum variance formulation, though is slightly pessimistic.  We
assume zero error caused by the reconstruction of pseudo-open-loop
slope measurements which would be typical of a \dm with closed-loop
feedback as is the case with the \eelt \glao \dm (M4).  A fitting step
is used to fit the reconstructed volume of turbulence onto the \dms.

\ignore{
pseudo-open-loop control so
that reconstruction is performed using open-loop wavefront sensor
information, even though the wavefront sensors operate in closed loop
behind the \glao \dm.  We are able to convert perfectly between
close-loop and pseudo-open-loop slope measurements, since we have an
ideal \dm, of which we know the exact mirror surface shape, meaning
that the uncertainties in \dm surface shape in a real system do not
affect our results.
}

\subsection{Investigations of LGS asterism radius}
It is known that analytical \ao modelling codes, for example Fourier
domain based codes, give optimistic performance estimates for
wide-field \ao systems \citep{miskaltao}, primarily due to the
assumption of infinite telescope diameter.  The optimum \lgs asterism
radius for EAGLE and other \moao systems is subject to some
uncertainly, and so here we investigate \ao performance as a function
of asterism radius.  Our results are also compared with those from an
analytical model (a Fourier domain code, \citet{2008JOSAA..26..219N}),
and with the \eso Octopus simulation code, as given by
\citet{miskaltao}.  The results presented here are also equally
applicable to \ltao systems due to the nature of the tomographic
problem.

\subsection{Investigations of actuator count}
Designs for \moao instruments such as EAGLE typically specify the
science channel \dms to have an actuator pitch equal to the \wfs
sub-aperture pitch.  For \eagle, this therefore corresponds to a
requirement for twenty $85\times85$ actuator \dms.  Current \dm
technologies have not been scaled to this many actuators, and
development of a suitable high-order \dm technology will introduce both
cost and risk to an \elt \moao instrument.  Here, we investigate the
impact that reducing \moao \dm actuator count has on \ao performance.
We take advantage of a \glao \dm, which has a pitch equal to that of
the \wfss, providing a global \ao correction.  Ground layer turbulence
is often strongest \citep{2010MNRAS.406.1405O}, so we hypothesise that
once a \glao correction has been applied, a reduced actuator count might
then be sufficient to perform \ao correction of the remaining turbulence
along the line of sight of each science object without significantly
reducing performance.  We also investigate the impact on \ao
performance if the \glao \dm actuator count is also reduced
simultaneously with the \moao \dm, for completeness.

To perform these investigations, our simulation consists of a \glao
\dm which is used to correct the tomographically estimated ground
layer turbulence, and a \moao \dm which corrects the higher layer
turbulence along the direction of the science object.

Here, we consider only the case where the \moao \dms are conjugated to
ground level.  However, \citet{basdenconf1} has previously
demonstrated the benefit of conjugating \moao \dms above ground level,
allowing a directional correction to be applied, widening the \moao
field of view by reducing anisoplanatism inside the \moao field.
Because we simulate only a single atmospheric layer at ground level,
our \glao correction may be optimistic, and a further study of this
effect is planned in future work.

\subsection{Investigations of mechanical stroke requirements}
The \dms used for science channel correction in a \moao system are
likely to have limited stroke, due to a combination of small physical
size and high actuator density.  A large number of such \dms are
required for an \moao instrument, and so the reduction in cost that
can be made by reducing \dm stroke requirement can be significant.  We
investigate the impact that reducing stroke will have on \ao
performance by considering two cases.  First that the \glao \dm has
unlimited stroke, whilst the \moao \dm has a restricted stroke.
Secondly, for completeness, we consider the case when all correction
is performed by the \moao \dms, and the impact that limited stroke
then has.  We simulate a restricted stroke by clipping \dm actuators
to the maximum allowed mechanical stroke.

\subsection{Investigations of DM misalignment}
The relative alignment between \wfss and \dms is critical for any \ao
system, and the tolerance to which the alignment between these
components must be maintained is an important design consideration.
We investigate the impact that misalignments have on \ao system
performance, including both lateral shifts and rotations.  To model
these effects, we shift or rotate the \dm surface once the \dm demands
have been applied to the mirror, and thus the corrected wavefront
contains the effects of these shifts and rotations.  Here, we do not
consider the \glao and \moao \dms separately, rather for simplicity,
we use only a \moao \dm (also correcting the ground layer), and shift
or rotate this.

\section{Results and discussion}
\subsection{LGS Asterism radius}
\begin{figure}
\includegraphics[width=\linewidth]{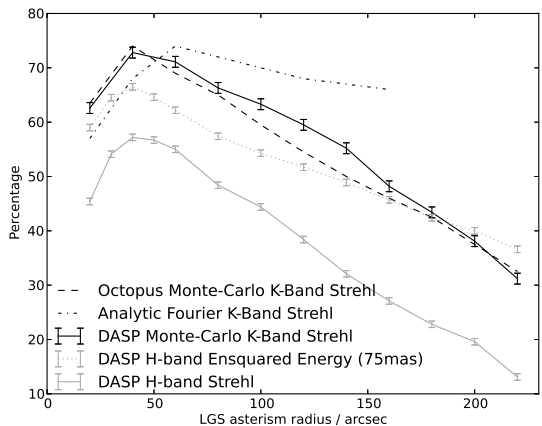}
\caption{A figure showing on-axis K-band Strehl ratio (black lines) as
  a function of LGS asterism radius.  The black solid curve presents
  the DASP results that we have obtained, the dashed curve is from the
  ESO Octopus Monte-Carlo simulation, and dot-dashed is from an
  analytic Fourier code.  For comparison with the remainder of this
  paper, H-band results are also shown in grey, with dotted grey being
  ensquared energy within 75~mas, and solid grey being Strehl ratio.}
\label{fig:asterism}
\end{figure}

\begin{figure}
\includegraphics[width=2.5cm,angle=270]{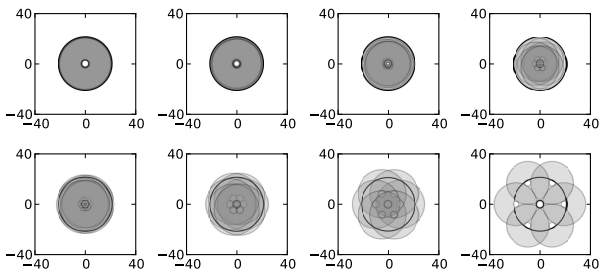}
\vspace{1cm}
\caption{A figure showing LGS overlap at different atmospheric heights
  for (top row) an asterism radius of 40~arcsec and (bottom row)
  an asterism radius of 220~arcsec.  The LGS beacon is focused
  at 90~km, and overlap at 2.25~km, 4.5~km, 9~km and 18~km is shown.
  The scale is in meters.}
\label{fig:ast2}
\end{figure}

An \ao system with multiple \ngss will always offer best on-axis
performance when the asterism radius is zero, i.e. the \scao case.
However, with \lgss this is not the case.  Due to the finite altitude
of the \lgss focal anisoplanatism is observed, thereby reducing \ao
performance. The effect of focal anisoplanatism can be reduced by
using multiple \lgss to sample a greater volume of atmosphere above
the telescope. Increasing the radius of the \lgs asterism past some
optimal diameter will however reduce \ao performance due to poor
sampling of higher altitude turbulence.  Fig.~\ref{fig:asterism} shows
simulation results as a function of asterism radius, comparing our
Monte-Carlo results with both Monte-Carlo results from another
independent Monte-Carlo code (\eso Octopus, \citet{miskaltao}), and an
analytic Fourier code \citep{2008JOSAA..26..219N}.  These simulations
all use parameters as closely matched as possible, including the same
atmospheric turbulence profiles, telescope diameter, guide star number
and \dm order.  These results are for K-band Strehl ratio
(2.2~$\mu$m).  As can be seen, the Monte-Carlo codes are in close
agreement.  The optimum asterism radius is shown to be about
40~arcsec, and Fig.~\ref{fig:ast2} shows the \lgs overlap for
different atmospheric heights at this diameter, showing that the
on-axis cone of turbulence is indeed well sampled except for at the
edges of the very highest layer.  The overlap for the nominal EAGLE
asterism radius of 220~arcsec is also shown, displaying reduced guide
star overlap, corresponding to poorer reconstruction of turbulence,
particularly at higher altitudes.  It should be noted that the
analytic code gives a different slope for the dependency of
performance on asterism radius, due to the infinite pupil assumption
\citep{miskaltao}.

\begin{figure}
\includegraphics[width=0.5\linewidth]{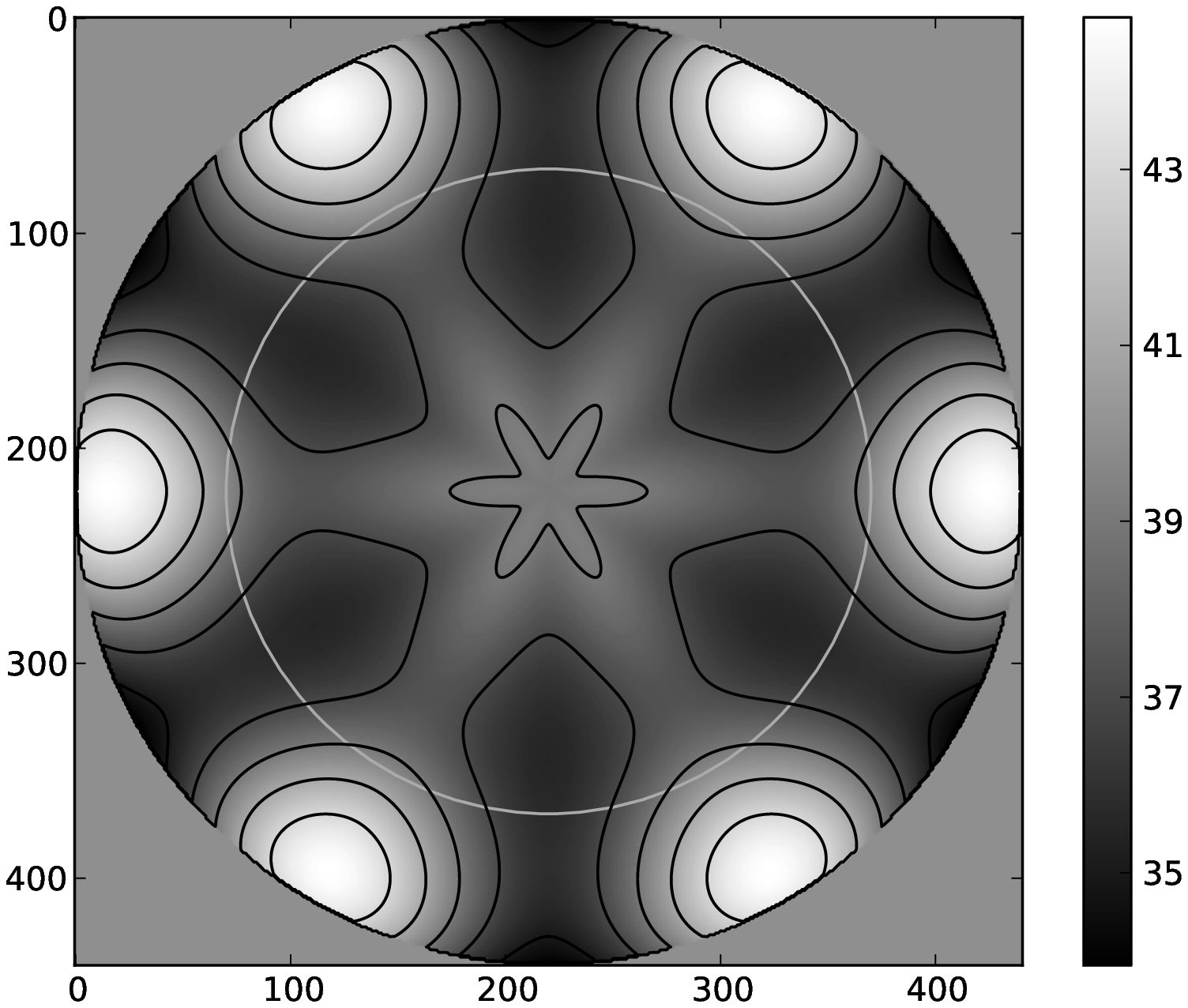}\includegraphics[width=0.5\linewidth]{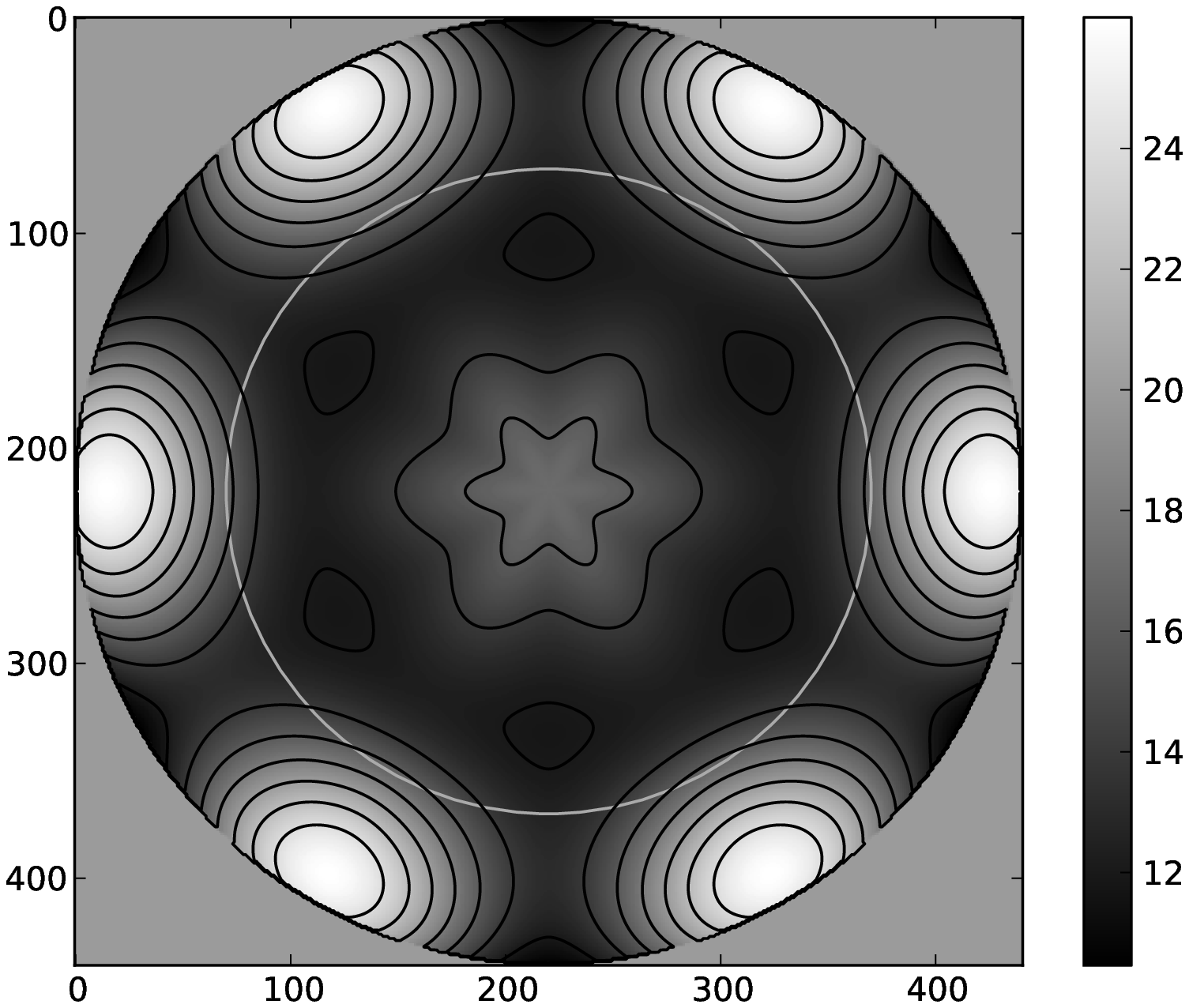}
\caption{(a) A figure showing H-band MOAO performance (ensquared energy in
  $75\times75$~mas) over the 440~arcsec technical field of view of an
  EAGLE-like instrument, with a LGS asterism diameter of 440~arcsec.
  The centred grey circle represents a 5~arminute field, and
  contours are spaced by 2\%, starting at 35\% ensquared H-band
  energy.  (b) As for (a), showing Strehl ratio over the field.}
\label{fig:astmap}
\end{figure}

Fig.~\ref{fig:astmap}(a) shows H-band ensquared energy within
$75\times75$~mas over the entire \eagle field of view, with a
220~arcsecond \lgs asterism radius.  Over the 5~arcminute science
field (represented by a grey circle), the variation in ensquared
energy ranges from 35--40\%.  Fig.~\ref{fig:astmap}(b) shows the
variation of Strehl ratio over this field.  Taking advantage of \ngss
available within the technical field of view would allow this
performance to be increased \citep{2010SPIE.7736E..25Rshort} though we
do not consider this further here.

Throughout the rest of this paper, results are given for H-band
ensquared energy within 75~mas, and for comparison purposes, these
results are also shown in Fig.~\ref{fig:asterism}.  The error bars in
this figure are calculated from the variance of multiple simulation
runs, and are at the sub 1~\% level for all further results presented
here, and thus are not shown for clarity.

\subsection{Actuator count}
\begin{figure}
\includegraphics[width=\linewidth]{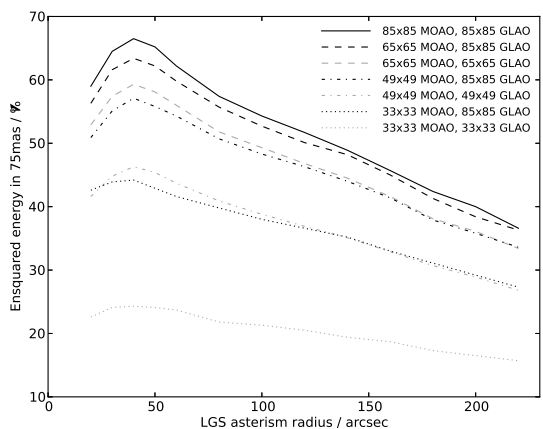}
\caption{A figure showing AO performance as a function of LGS asterism
  radius for the different DM actuator counts given in the legend.  A
  ground layer (GLAO) correction followed by an MOAO correction (with
  no ground layer component) is performed.  Stroke is unlimited for
  both DMs.  It should be noted that cases with equal order for both
  GLAO and MOAO DMs are identical to a case using only an MOAO DM of
  the same order that includes ground layer correction and is not
  stroke limited.}
\label{fig:nact}
\end{figure}

It can be seen from Fig.~\ref{fig:nact} that when a \glao \dm is
present, the actuator density for \moao \dms (which perform only
higher layer correction) can be relaxed somewhat without dramatically
affecting \ao performance.  This is an encouraging result for \moao
system designers, because it allows what is a high cost single
component (the \dm), of which many (20 for \eagle) are required for an
\moao system, to have its specification reduced.  Additionally, this
reduces the computational demands (which typically scale as the square
of the total number of actuators) placed on the necessary real-time
control system \citep{basden9} (nearly a factor of three reduction in
computational requirements moving from $85\times85$ to $65\time65$
actuators).  For \eagle this is important, because although it has
been shown that real-time control on this scale is a tractable problem
\citep{basden11}, reducing computational demands in wavefront
reconstruction will provide the opportunity for additional algorithms
to be used to further improve \ao system performance, such as the
brightest pixel selection algorithm \citep{basden10} successfully
demonstrated with \canary.

The presence of a high order \glao \dm is, unsurprisingly, helpful,
compared with the case where both \dms are of equal, lower, order
(which is identical to using only a \moao \dm of
this order, also including ground layer correction, in these
simulations).  This is because using a higher order \glao \dm reduces
the fitting error of the \dm to ground layer turbulence (which
increases with actuator pitch).  In particular, if the \moao \dm order
is dropped to $33\times33$ actuators, which represents a readily
available \dm, \ao performance is doubled when the \glao \dm is
present at full $85\times85$ actuators (\eelt M4 scale) compared to
when this is also dropped to $33\times33$ actuators (which is also
equivalent to not using a \glao \dm, and using the \moao \dm to
perform all correction including the ground layer).  We can see from
Fig.~\ref{fig:nact} that there is only a small performance loss of
about 2--3\% when using a $65\times65$ actuator \moao \dm compared
with a $85\times85$ actuator \dm when the \glao \dm is present.  It
should be noted that we have used these \dm actuator counts for ease
of simulation, and that removing one row and column to match currently
available \dms (for example the $64\times64$ and $32\times32$ actuator
\dms available from Boston Micro Machines) will have little impact on
performance, as can be seen from the trend of performance with
actuator count.

\subsubsection{Pseudo-open-loop control considerations}
Changes applied to the \glao \dm on the \eelt are sensed by the \wfss,
and thus it is necessary to operate using a pseudo-open-loop
controller so that minimum variance wavefront reconstruction can be
performed.  There will always be some uncertainty in the mirror
surface shape, however small, and this will lead to a non-zero
pseudo-open-loop error, though this will be minimised by accurate \dm
surface position sensors (either optical or mechanical).  It is
interesting to consider the impact of this error source here.  

If the \moao \dm actuator count is to be constrained for cost reasons,
we have the choice of either using it alone, or in conjunction with
the higher order \glao \dm accepting the additional pseudo-open-loop error.
\dm fitting error in radians squared is given approximately by
(\citet{hardy}, p196):
\begin{equation}
\sigma_{\mathrm F}^2 \approx f \left( \frac{d}{r_{\mathrm 0}}\right)^{\left(
  \frac{5}{3} \right)}
\end{equation}
where $f$ is a constant that depends on the \dm (typically around 0.28), $d$
is the actuator pitch and $r_{\mathrm 0}$ is Fried's parameter.

Considering only the effect of ground layer turbulence, at 1650~nm,
this gives a fitting error contribution of about 57~nm for an
$85\times85$ actuator \dm, 71~nm for a $65\times65$ actuator \dm,
91~nm for a $49\times49$ actuator \dm and 127~nm for a $33\times33$
actuator \dm.  Therefore if we perform ground layer correction with a
closed-loop $85\times85$ actuator \dm, we can accept up to 42~nm
pseudo-open-loop error, and still obtain better performance than if
using an open-loop $65\times65$ actuator \dm (by adding error terms in
quadrature).  Likewise, we can accept up to 71~nm pseudo-open-loop
error before it is better to use a $49\times49$ actuator open-loop
\dm, and up to 113~nm error before it is better to use a $33\times33$
actuator open-loop \dm.  The actual pseudo-open-loop error for the
\eelt M4 \dm is not yet known, however one would hope that it would be
below these levels due to accurate position sensors.

\ignore{
Calculations are:

numpy.sqrt(.28*(42./84./(.1993*(1650./500)**(5./3)))**(5./3))*(1650/2./numpy.pi)

where 0.1993 = 0.135 * (0.5224)**(-3/5.)

Change 84 to 64, 48 or 32 for the other values.

}

\subsection{DM stroke requirements}
\begin{figure}
\includegraphics[width=\linewidth]{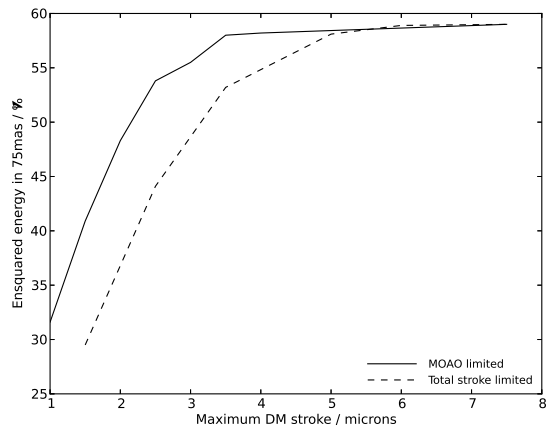}
\caption{A figure showing AO performance as a function of maximum MOAO
  DM stroke (solid) for a LGS asterism with 20~arcsec radius.
  Also shown (dashed) is performance when total stroke is limited
  (i.e. assuming the GLAO correction is not present).}
\label{fig:stroke}
\end{figure}

\begin{figure}
\includegraphics[width=\linewidth]{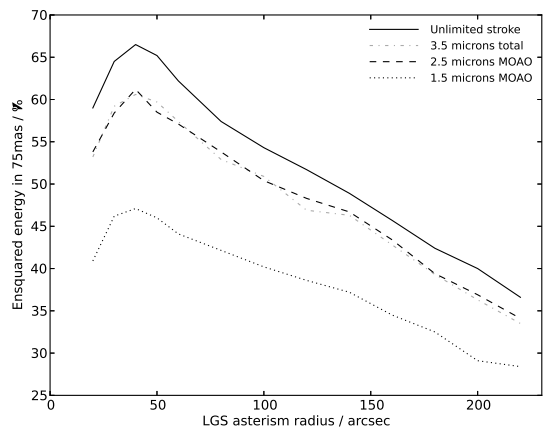}
\caption{A figure showing AO performance as a function of LGS asterism
  radius, for an MOAO DM with unlimited stroke (solid), a maximum stroke of
  2.5~$\mu$m (dashed), and a maximum stroke of 1.5~$\mu$m (dotted).}
\label{fig:strokeAsterism}
\end{figure}

The \eelt contains a large \dm as part of the telescope optical train
conjugated close to the ground.  This \dm is physically large, and is
expected to have large (essentially unlimited) stoke.  The \moao \dms,
of which a large number is required, are likely to have limited
stroke, being physically small.  Fig.~\ref{fig:stroke} shows the
impact of maximum \moao \dm mechanical stroke on \ao performance for
the case of the narrowest asterism considered here (20~arcsec radius).
In this case, the \glao \dm is assumed not to be stoke limited.
Fig.~\ref{fig:strokeAsterism} shows \ao performance as a function of
\lgs asterism radius for the case of unlimited \moao \dm stroke, and
with stroke limited to 1.5~$\mu$m and 2.5~$\mu$m.  It is also
interesting to see how performance is affected when total stroke is
limited, and this is shown in Fig.~\ref{fig:stroke} and in
Fig.~\ref{fig:strokeAsterism} as a function of asterism radius.  It
should be noted that in this case, we assume only one \dm, rather than
two \dms each with limited stroke, i.e. operation without the \glao
\dm.

From Fig.~\ref{fig:stroke}, we can see that a maximum mechanical stroke of
3.5~$\mu$m for the \moao \dms will reduce performance by only a fraction
of a percentage point when compared with a stroke-unlimited \dm.  A
maximum stoke of 2.5~$\mu$m will lead to a slight reduction in
performance, while limiting stroke to 1.5~$\mu$m reduces performance by
a third.  It is interesting to note that the presence of the \glao \dm
allows the stroke requirements on the \moao \dms to be relaxed by
about 1~$\mu$m (Fig.~\ref{fig:stroke}).

\subsection{DM misalignments}
During \ao system calibration, the relative alignment of the \wfss to
the \dm actuators is generally encoded within the system using a
control matrix or other means. Any unobserved deviation in position
between the \wfss and \dm after the calibration procedure can result in
a reduction in system performance.

\begin{figure}
\includegraphics[width=\linewidth]{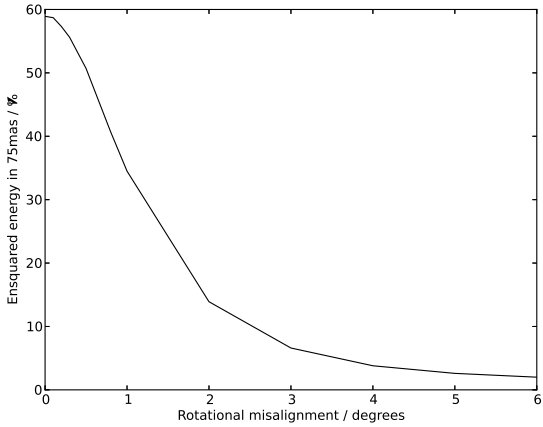}
\caption{A figure showing AO performance as a function of rotational
  DM misalignment for a 20~arcsec LGS asterism radius (without the
  GLAO DM).}
\label{fig:rotation}
\end{figure}

\begin{figure}
\includegraphics[width=\linewidth]{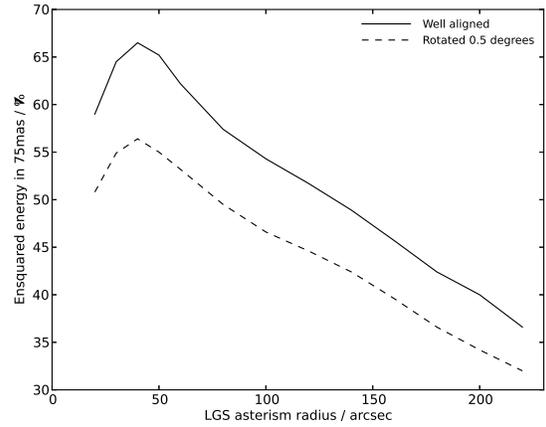}
\caption{A figure showing AO performance as a function of LGS asterism
  radius for rotational DM misalignments of 0$^\circ$ (i.e.\ well
  aligned) and 0.5$^\circ$.}
\label{fig:rotationb}
\end{figure}

The rotation of a \dm relative to the expected position (and thus the
position for which \dm demands are computed) affects performance as
shown in Fig.~\ref{fig:rotation}.  It is clear here that performance
is affected even for small rotation angles, falling steeply for angles
larger than 0.2$^\circ$.  This angle corresponds to a shift of about
14\% of a sub-aperture for the outer ring of sub-apertures, which is a
2.3 pixel shift in actuator position relative to the \wfs
sub-aperture.  Fig.~\ref{fig:rotationb} shows \ao performance as a
function of asterism radius when the relative \dm rotation is
0.5$^\circ$, with unrotated performance shown for comparison, showing
an effectively constant drop in performance when misalignment
occurs.

\begin{figure}
\includegraphics[width=\linewidth]{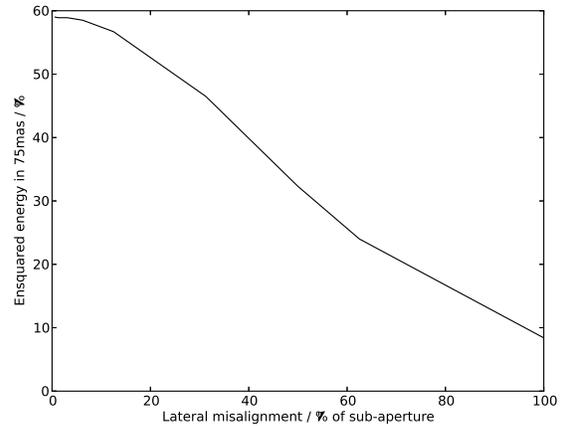}
\caption{A figure showing AO performance as a function of lateral
  DM misalignment as a percentage of a sub-aperture (with $16\times16$ pixels
  per sub-aperture), for a 20~arcsec LGS asterism radius.}
\label{fig:align}
\end{figure}

\begin{figure}
\includegraphics[width=\linewidth]{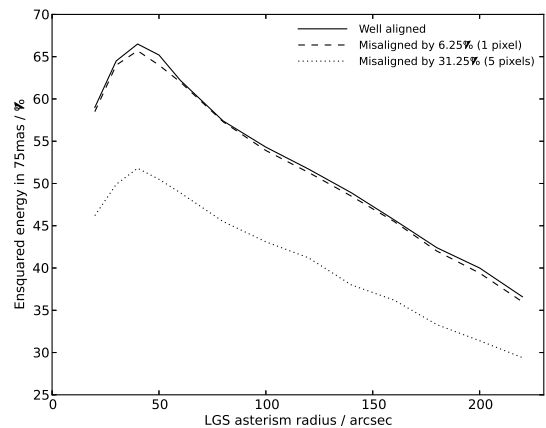}
\caption{A figure showing AO performance as a function of LGS asterism
  radius for lateral DM misalignments of 0 (i.e.\ well aligned), 1 and
  5 pixels.}
\label{fig:alignb}
\end{figure}

The relative lateral shift of a \dm between its assumed and actual position
affects performance as shown in Fig.~\ref{fig:align}.  Here, we can
see that shifts of up to 2 pixels (12.5\% of a sub-aperture) lead to
only small drops in performance, while for larger misalignments, \ao
performance begins to fall more rapidly.  Fig~\ref{fig:alignb} shows
\ao performance as a function of asterism radius for misalignments
of one and five pixels, as well as the well aligned case for comparison.

\section{Conclusions}
We have investigated the impact of \lgs asterism radius and various
\dm characteristics, including \dm order, maximum mechanical \dm
stroke, and \dm alignment tolerances for an \elt scale \moao system,
using a full end-to-end Monte-Carlo \ao simulation tool.  These
simulations have been based on the conceptual designs for the \eagle
\moao instrument.

We have not sought to give definitive answers to the questions
investigated here, rather specifying how \ao performance is affected
by these parameters.  This information can then be used during the
design and specification of the relevant instrument.  For example, we
have shown that small misalignments of the \dm lead to only small
drops in \ao performance.  The performance trade-off decisions that
can be made are left to science instrument considerations.  However,
from our study, it is helpful to make some observations as follows.

\Ao performance as a function of \lgs guide star asterism radius has
been shown to fall more steeply than suggested by Fourier based
analytical codes which assume infinite pupil diameters, confirming
previous results \citep{miskaltao}.  However, we have shown that for a
perfect \moao system on a 42~m \elt, with six \lgss placed in a ring
with an asterism radius of 220~arcsec and with perfect tip-tilt
anisoplanatism correction, H-band performance across the science field
of view is sufficient to give more than 35\% ensquared energy within
75~mas of the resulting science \psf, which is better than the
requirement for the \eagle \moao \elt instrument.

We have shown that the presence of a high order global \glao \dm
allows the requirements for \moao \dms to be reduced.  A $49\times49$
actuator \dm meets the \eagle on-axis \ao performance
requirement with the largest \lgs asterism radius, if the assumptions
made here are valid (namely no misalignment, and perfect tip-tilt
correction).  This will allow the cost of \eagle to be greatly
reduced.  Using a $65\times65$ actuator \dm gives almost no reduction
in \ao performance compared with a full $85\times85$ actuator \dm
matched to the \wfs sub-aperture count.

To preserve \ao performance the \moao \dms must have a maximum stroke
capability of at least 2.5~$\mu$m, with 3.5~$\mu$m being a goal.

\Dm to \wfs alignment tolerances must be kept to within an
eighth of a sub-aperture of calibrated position, so that performance
is not significantly affected.  For rotation, this represents an angle
of 0.2$^\circ$ being the maximum misalignment from expected \dm
position to avoid significant performance reductions.

In this study, the effect of \ngss hasn't been included meaning that
these results are slightly pessimistic, and neither have variations in
sodium layer, or telescope vibrations.  These issues will be the
subject of future work.

\section*{Acknowledgements}
The authors would like to thanks M.~Le~Louarn and G.~Rousset for
providing advice and comments, and for the referees who helped to
improve this paper.

\bibliographystyle{mn2e}

\bibliography{mybib}
\bsp

\end{document}